SCSE_65_2018

# Improved hierarchical role based access control model for cloud computing

* N.N.Thilakarathne and Dilani Wickramaaarachchi

*Department of Industrial Management, University of Kelaniy, Srilanka*

*\*Neranjanthi@gmail.com*

## Abstract

Cloud computing is considered as the one of the most dominant paradigm in field of information technology which offers on demand cost effective services such as Software as a service (SAAS), Infrastructure as a service (IAAS) and Platform as a service (PAAS).Promising all these services as it is, this cloud computing paradigm still associates number of challenges such as data security, abuse of cloud services, malicious insider and cyber-attacks. Among all these security requirements of cloud computing access control is the one of the fundamental requirement in order to avoid unauthorized access to a system and organizational assets. Main purpose of this research is to review the existing methods of cloud access control models and their variants pros and cons and to identify further related research directions for developing an improved access control model for public cloud data storage. We have presented detailed access control requirement analysis for cloud computing and have identified important gaps, which are not fulfilled by conventional access control models. As the outcome of the study we have come up with an improved access control model with hybrid cryptographic schema and hybrid cloud architecture and practical implementation of it. We have tested our model for security implications, performance, functionality and data integrity to prove the validity. We have used AES and RSA cryptographic algorithms to implement the cryptographic schema and used public and private cloud to enforce our access control security and reliability. By validating and testing we have proved that our model can withstand against most of the cyber-attacks in real cloud environment. Hence it has improved capabilities compared with other previous access control models that we have reviewed through literature.

## Introduction

Cloud is one of the major and dominate technology that pave the way for digital transformation across the globe. It is a model for providing convenient on demand network access for computing resources such as applications, services, servers and storages that can be rapidly provisioned and released with minimal management effort or service provider interaction (Bibin, 2013).Cloud has a lot of advantages mainly in ubiquitous services where everybody can access computer services through internet. This cloud model composed of three service delivery models mainly SAAS, PAAS, and IAAS. Depending on the type of data that you are working with, cloud computing come in three forms. Public cloud, Private cloud and the Hybrid cloud. Along with the rapid steady development of the cloud applications cloud computing cyberattacks are also increased and cloud itself create a good attacking surface for hackers. (Faisal at el., 2015)Denial of Service attacks (DOS attacks), Authentication attacks, Side channel attacks, Cryptographic attacks ,and Inside Job attacks are best attack vectors for those hackers and due to these generalized attacks we need a better security reinforcement for cloud computing as it can lead to a major cyber-attack. Due to this reason there are number of security challenges associated with utilizing cloud computing such as data security, abuse of cloud services, malicious insider and cyber-attacks. (Aditya and Ashish, 2013)

Among all security requirements of cloud, access control is one of the fundamental requirements in order to avoid unauthorized access to system and organizational assets and cloud access control models can be traditionally categorized into Discretionary access control (MAC), Mandatory access control (DAC) and Role Based Access Control (RBAC).In the Discretionary access model (DAC) model ,the administrator of the object decides its access permissions for users based on an access control list and in the Mandatory access control (MAC) model access permissions are decided by the administrator of the system. In the Role-based access control model (RBAC), a user has access to a resource based on his/her assigned role in the system. Younis at el.,2014)Roles are defined based on job functions and permissions are defined on authority and responsibilities of the job. Operations on the resources are invoked based on the permissions. RBAC models are more scalable than the discretionary and mandatory access control models, and more suitable for use in cloud computing environments (Natarajan, 2011).

## Methodology

A thorough Literature Survey has been done to find out the potential gaps and the features that are not available in the existing access control mechanisms. Based on that we have identified Hierarchical RBAC access control model to be improved with hybrid cloud architecture *(combination of public and the private cloud)* and hybrid encryption schema (*AES -128 bit symmetric key algorithm and RAS 1024 bit public key encryption*).Our proposed model uses AES for encrypting and decrypting cloud data and RSA public key encryption algorithm is used to encrypt the secret key generated by the AES cryptographic algorithm.

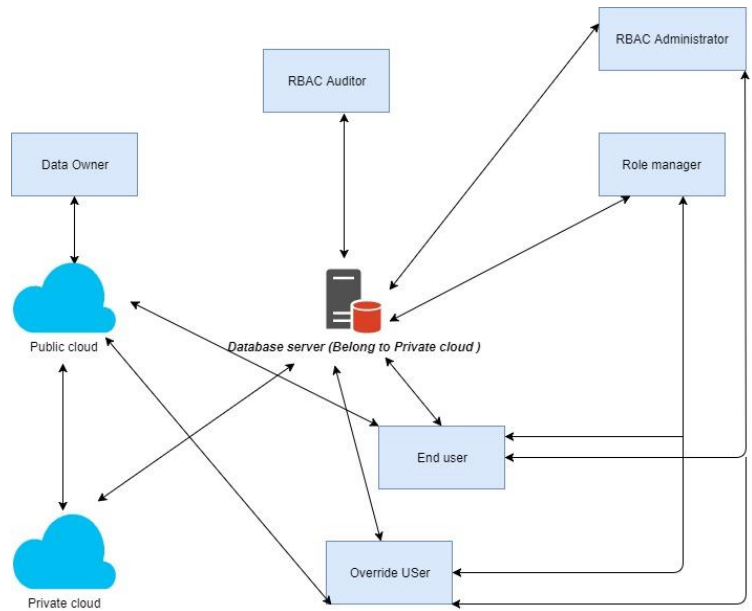

Figure 1: Architecture of our proposed access control model

As depicted in figure 1, proposed architecture of the access control model comprised of Public Cloud, Private Cloud, System Administrator, Role Manager, Data Owner, End User, and Override User.

Public information are encrypted using symmetric AES algorithm which will be saved in the public cloud. The organizations stores only critical and confidential information in private cloud. Private cloud is built on an internal data center that is hosted and operated inside the organization. The amount of information stored in private cloud is relatively smaller when compared to public cloud. Administrator of role based system provides authentication to end users. End users are just to access data from cloud. They cannot do any modifications or updates to original data. They cannot communicate directly with the private cloud as they do not possess access permission. Administrators generate the system parameters. System parameters represent the position of the role and stored that role in private cloud. Also administrators manages role hierarchy. Role manager will be there for manage roles for users. According to the role and authorization from the role manager, user gets access permission to cloud data. The access policies related to authorization and roles are stored in private cloud.

**Results**

We have implemented the proposed model in ASP.NET, C# using Visual Studio 2017 IDE and for the public cloud we have taken an instance from Microsoft azure. Testing and validation done on Microsoft Azure instance that has single core, 1.75G Ram with 10 GB storage. For the implementation on the private cloud we have chosen a SQL server instance from the Microsoft Azure with inbuilt firewall with 50 MB storage. (SQL Server database as a service). For the public cloud we have deployed our developed model on to the Microsoft Azure as an app service.

After the successful implementation, we have tested our model for performance, data integrity, security implications and functional analysis. From the performance testing we conclude that encryption time took more time than decryption time and we can see when the file size increases both the encryption and decryption time are gradually increasing. For the Data integrity testing we have compared and verified MD5, SHA-1, SHA-256 and SHA-512 values for both original file and decrypted download file and both types of files have same values when testing for data integrity. Thus we conclude that the file integrity of resources was preserved during encrypting and decrypting which enforce the reliability and the integrity of our access control model.

We have performed three vulnerability assessment tests to compare and validate our model in real cloud environment and results from these three tests have showcased that our access control model withstand against most of the cyber-attacks. Further for functional testing we have compared our access control model with conventional access control models for cloud, based upon the functionality that our model and those models can offer as depicted in figure 2.

| NO | Comparison criterion | DAC | MAC | RBAC | ABAC | Our Model |
|---|---|---|---|---|---|---|
| 1 | Least privilege principle | N | N | Y | Y | Y |
| 2 | Separation of duties | N | N | Y | Y | Y |
| 3 | Scalability | N | N | Y | N/A | Y |
| 4 | Auditing | Y | Y | Y | Y | Y |
| 5 | Policy management | Y | N | Y | Y | Y |
| 6 | Flexibilities of configuration | N | N | Y | Y | Y |
| 7 | Delegation of capabilities | Y | N | N | N | Y |
| 8 | Hybrid cloud architecture | N | N | N | N | Y |
| 9 | Role hierarchy management | N | N | Y | Y | Y |
| 10 | Operational and situational awareness | N | N | N | N | Y |

Y-Yes, N-No, N/A-Not Applicable

Figure 2: Functional analysis and testing

## Conclusion

The main objective of this research was to come up with an improved access control model, that can be utilize in public cloud and also that can be utilize for secure cloud data storage. Before proposing our model we have reviewed almost every access control model in cloud and none of the researches were targeted about the security implications of access control in real cloud computing environment. We have showed the experimental result of our implemented access control model through the perspective of performance, data integrity and security in real cloud environment. We observed that time taken for encryption and decryption is efficient on public cloud and maintaining data storage in public cloud is efficient as it is highly scalable, cost effective and provides redundancy for the organization data. Further we observed that original and decrypted data after encryption was same, when data integrity checking thus it will enhance the reliability and assurance of our model. From the security perspective we have observed that it's not vulnerable to exploit and hence it provide the reliability and security for our access control model.

We believe that the proposed model is useful in various commercial and non – commercial situations as it implements the hierarchical cryptographic role based access control policies based on the job functionality and user requests in an organization for providing secure data storage in the real cloud environment enforcing hybrid cryptographic techniques for providing security for underlying data along with hybrid cloud architecture.